\documentclass[aps,prd,twocolumn,floatfix,preprintnumbers,altaffilletter,superscriptaddress,nofootinbib]{revtex4-1}
\usepackage{graphicx,url,amssymb,amsmath,longtable,rotating,color,units,wasysym,subfigure,epsfig,multirow,epstopdf,microtype}
\usepackage[dvipsnames]{xcolor}
\usepackage[colorlinks,urlcolor=NavyBlue,citecolor=NavyBlue,linkcolor=NavyBlue]{hyperref}
\usepackage[bottom]{footmisc}
\usepackage[]{comment}

\newcommand{\evidence}{Z}
\newcommand{\hyp}[1]{{\cal H}_{\rm #1}}

\newcommand{\eq}[1]{Eq.\ \eqref{e:#1}}
\newcommand{\fig}[1]{Fig.~\ref{fig:#1}}

\newcommand{\heading}[1]{\section{#1}}

\newcommand{\ntrigs}{983}
\newcommand{\ninjs}{648}

\newcommand{\dcc}{LIGO-P1700414}

\begin{document}

\title{Enhancing confidence in the detection of gravitational waves\\from compact binaries using signal coherence}

\author{Maximiliano Isi}
\email[]{misi@ligo.caltech.edu}
\affiliation{%
LIGO, California Institute of Technology, Pasadena, CA 91125, USA
}
\author{Rory Smith}
\email[]{rory.smith@ligo.org}
\affiliation{%
LIGO, California Institute of Technology, Pasadena, CA 91125, USA
}
\affiliation{Monash Centre for Astrophysics, School of Physics and Astronomy, Monash University, VIC 3800, Australia}
\affiliation{OzGrav: The ARC Centre of Excellence for Gravitational-Wave Discovery, Monash University, VIC 3800, Australia}

\author{Salvatore Vitale}
\email[]{salvatore.vitale@ligo.mit.edu}
\affiliation{
LIGO, Massachusetts Institute of Technology, Cambridge, Massachusetts 02139, USA
}%

\author{T.\ J.\ Massinger}
\affiliation{%
LIGO, California Institute of Technology, Pasadena, CA 91125, USA
}

\author{Jonah Kanner}
\affiliation{%
LIGO, California Institute of Technology, Pasadena, CA 91125, USA
}

\author{Avi Vajpeyi}
\affiliation{%
Physics Department, The College of Wooster, Wooster, Ohio 44691, USA}
\affiliation{%
LIGO, California Institute of Technology, Pasadena, CA 91125, USA
}

\date{\today}

\begin{abstract}
We show that gravitational-wave signals from compact binary mergers may be better distinguished from instrumental noise transients by using Bayesian models that look for signal coherence across a detector network.
This can be achieved even when the signal power is below the usual threshold for detection.
This method could reject the vast majority of noise transients, and therefore increase sensitivity to weak gravitational waves.
We demonstrate this using simulated signals, as well as data for GW150914 and LVT151012.
Finally, we explore ways of incorporating our method into existing Advanced LIGO and Virgo searches to make them significantly more powerful.
\end{abstract}

\pacs{Valid PACS appear here}
\preprint{\dcc}

\maketitle


\heading{\label{sec:intro}Introduction}
A pair of neutron stars or black holes merges somewhere in the observable universe roughly every 15--200s, releasing large amounts of energy in the form of gravitational waves (GWs) \cite{GW150914_paper,GW151226-DETECTION,O1:BBH,GW170104_paper,GW170814,GW170817,GW170817_stoch}.
One of the limiting factors in detecting such GWs with existing detectors, like Advanced LIGO (aLIGO) and Virgo \cite{TheLIGOScientific:2014jea,TheVirgo:2014hva}, is data contamination by instrumental noise transients (\emph{glitches}) that may mimic astrophysical signals \cite{GW150914_detchar}.
Glitches can lower the inferred statistical significance of GW signals, making their detection more difficult.
In this paper, we show how signal coherence may be used to address this problem by significantly improving our ability to distinguish genuine GW signals from glitches using Bayesian model comparison.

In particular, we demonstrate that Bayesian models---as proposed in \cite{VV2010}---may successfully distinguish real GWs from glitches by using the fact that the former must be {\em coherent} across detectors, while the latter will generally not be.
Here, coherence means that a real GW must produce strain signals in different instruments that: $(i)$ are coincident in time (up to a time-of-flight delay); $(ii)$ are well-described by a compact-binary-coalescence (CBC) waveform; and $(iii)$ share a phase evolution consistent with a single astrophysical source.
In contrast, glitches should not be expected to fully satisfy these criteria.
Making full use of this information---the expected coherence of signals and incoherence of glitches---may allow us to detect weaker signals than is currently possible.

From a subset of glitches and detection candidates ({\em triggers}) from aLIGO's first observation run (O1), we find that: $(a)$ the majority of glitches are markedly more incoherent than coherent across detectors, irrespective of their loudness or the detection significance assigned by one of the main detection pipelines; $(b)$ simulated signals can be identified by their coherence, as long as they are distinguishable from Gaussian noise in at least two detectors; and finally, $(c)$ the ``gold-plated'' detection GW150914 (detection significance $> 5.1\sigma$)  \cite{GW150914_paper} and the ``silver-plated'' \textit{candidate} LVT151012 (detection significance $\sim 2.1\sigma$) \cite{O1:BBH} are both decidedly more coherent than incoherent.
This study of real data thus implies that the Bayesian comparison of coherent and incoherent signal models has the potential to significantly improve the sensitivity of CBC searches, even with currently available computational resources.

\heading{\label{sec:searches} Searches}
Templated searches for transient gravitational waves work by constructing a ranking statistic based on matched filtering \cite{Cannon2011Early,Cannon2013FAR,gstlal-methods,Nitz:2017svb,Usman:2015kfa,Canton:2014ena}.
In principle, to make a rigorous statement about the statistical significance of a pair of time-coincident triggers, it is necessary to know the probability that a given event was produced by instrumental noise, rather than an actual GW.
This likelihood may be estimated empirically from the value of the ranking statistic for a large representative set of triggers known with certainty to be spurious.
Such a set of signal-free triggers is denoted {\em background}, in contrast to the {\em foreground} of candidates that may contain a signal.

Because detectors cannot be physically shielded from gravitational waves, \textit{ad hoc} data analysis techniques must be used to estimate the background.
One such strategy is to construct \textit{time slides} by applying relative time offsets (longer than the light-travel time between sites) between the data of different detectors \cite{Usman:2015kfa,Canton:2014ena}.
Detection significance can then be inferred, in a frequentist way, by comparing the value of the ranking statistic for a time-coincident foreground trigger to that of time-slid background triggers.
The rate at which background triggers are produced with a given value of the ranking statistic is usually referred to as the {\em false-alarm rate} (FAR).

Efficient signal detection requires a ranking statistic that extracts the most information from the data, in order to discriminate between noise and weak astrophysical signals.
However, existing CBC searches are {\em not} optimal in this sense: they do not incorporate knowledge of {\em all} features that may distinguish GWs from noise.
Moving towards an optimal statistic is a great challenge, but one large step is to demand that foreground triggers in two or more detectors should be {\em better} described as coherent gravitational-wave signals, rather than incoherent glitches.
Importantly, it is not enough to provide some measure of coherence: one must also prove that an incoherent model is not {\em more} successful at describing the data.

\heading{\label{sec:bcr}Coherence vs incoherence}
To achieve this, we introduce the {\em Bayesian coherence ratio} (BCR):
the odds between the hypothesis that the data comprise a coherent CBC signal in Gaussian noise ($\hyp{S}$), and the hypothesis that they instead comprise incoherent instrumental features ($\hyp{I}$)---meaning each detector has \textit{either} a glitch in Gaussian noise ($\hyp{G}$), \textit{or} pure Gaussian noise ($\hyp{N}$).
For a network of $D$ detectors:
\begin{equation} \label{e:bcr}
\text{BCR} \equiv \frac{\alpha \evidence^{S}}{\prod^D_{i=1} \left[\beta\evidence^{G}_{i} + (1-\beta) \evidence^{N}_{i} \right]}\, ,
\end{equation}
where $\evidence^S$ is the evidence for $\hyp{S}$, and  $\evidence^{\rm G}_i$ and $\evidence^{\rm N}_i$ are, respectively, the evidences for ${\hyp{G}}_i$ and ${\hyp{N}}_i$ in the $i$\textsuperscript{th} detector.
The arbitrary weights $\alpha$ and $\beta$ parametrize our prior belief in each model: $\alpha = P(\hyp{S})/P(\hyp{I})$ and $\beta = P({\hyp{G}}_i \mid \hyp{I}) = 1 - P({\hyp{N}}_i \mid \hyp{I})$ for all $i$ (see, e.g., Eq.~(59) in \cite{Isi2017}).
These priors will be chosen to minimize overlap between the signal and noise trigger populations; their importance is studied in detail in Appendix \ref{app:weights}.

Evidences (marginalized likelihoods) are the conditional probability ($P$) of observing some data (${\bf d}_i$, for detector $i$) given some hypothesis ($\hyp{}$). 
For the coherent-signal hypothesis this is
\begin{align} \label{e:Z}
\evidence^{S} &\equiv P(\{{\bf d}_{i}\}_{i=1}^D \mid \hyp{S}) \\
&=\int p(\vec{\theta} \mid \hyp{S})\, p(\{{\bf d}_{i}\}_{i=1}^D\mid\vec{\theta}, \hyp{S})\, {\rm d}\vec{\theta}\, . \nonumber
\end{align}
The vector $\vec{\theta}$ represents a point in the space of parameters that describe the CBC signal, such as the component masses and spins; the terms in the integrand are the prior, $p(\vec{\theta}\mid \hyp{S})$, and the multi-detector likelihood, $p(\{{\bf d}_{i}\}_{i=1}^D\mid\vec{\theta}, \hyp{S}) = \prod_{i=1}^D p({\bf d}_{i}\mid\vec{\theta}, \hyp{S})$.
The specific functional form of the single-detector likelihood, $p({\bf d}_{i}\mid\vec{\theta})$, is derived from the statistical properties of the noise (e.g.\ a normal distribution for a Gaussian process).
The integral is performed numerically using algorithms like {\em nested sampling} \cite{Skilling:2006,Veitch:2014wba}.
In our case, the data ${\bf d}_i$ are the calibrated Fourier-domain output of each detector, but could generally be any sufficient statistic produced from it.

Because of their inherently unpredictable nature, it is impossible to produce a template that {\em a priori} captures all features of a glitch.
Therefore, we define a surrogate glitch hypothesis by the presence of simultaneous, but incoherent, CBC-like signals in different detectors.
Thus, for the $i$\textsuperscript{th} detector, the glitch evidence is 
\begin{align} \label{e:zg}
\evidence^{G}_i &\equiv P({\bf d}_i \mid \hyp{G}) \\
&= \int p(\vec{\theta}_i\mid \hyp{G})\, p({\bf d}_i\mid \vec{\theta_i}, \hyp{G})\, {\rm d}\vec{\theta_i}\, , \nonumber
\end{align}
where now we allow for a different set of signal parameters $\vec{\theta}_i$ at each detector.%
\footnote{Note that $\hyp{S}$ and $\hyp{I}$ are disjoint even if we do not explicitly exclude points from the parameter space satisfying $\vec{\theta}_i = \vec{\theta}_j$ for all $i \neq j$, because this condition defines a subspace that offers infinitesimal support to the prior in $\hyp{I}$ (see \cite{Li:2011cg,Isi2017}, or more general discussions in Ch.\ 4 in \cite{Sivia2006} or Ch.\ 28 in \cite{MacKay2005}).}
We will set $p(\vec{\theta}_i\mid \hyp{G})= p(\vec{\theta}_i\mid \hyp{S})$ and $p({\bf d}_i\mid \vec{\theta_i}, \hyp{G}) = p({\bf d}_i\mid \vec{\theta_i}, \hyp{S})$, but this may be relaxed to better capture specific glitch features, if necessary.
The surrogate $\hyp{G}$ model captures the portion of glitches that lie within the manifold of CBC signals and, in a sense, corresponds to the worst possible glitch---one that looks exactly like coincident CBC signals.
Variations of this strategy have been used before in the analysis of compact binary coalescences \cite{VV2010}, minimally-modeled transients \cite{Lynch:2015yin,Cornish:2014kda,Powell:2017gbj}, and continuous waves \cite{Keitel:2013wga,1742-6596-363-1-012041,Abbott:2017ylp}.
Other searches also make use of likelihood ratios in the detection process, but they do not rely on signal coherence (e.g.\ \cite{Cannon2013FAR,gstlal-methods}).

Finally, because we assume a perfect measurement of the detector noise power-spectral-density (PSD), the Gaussian-noise evidence is just the usual null likelihood.
For our Fourier-domain data, this is just
\begin{equation}
\evidence^{N}_i \equiv P({\bf d}_i \mid \hyp{N}) = \mathcal{N}({\bf d}_i)\, ,
\end{equation} 
where $\mathcal{N}({\bf d}_i)$ is a multidimensional normal distribution with zero mean and variance derived from the noise PSD \cite{Veitch:2014wba}.
In principle, this could be easily generalized to marginalize over poorly-known PSD parameters if needed.

\heading{\label{sec:analysis}Analysis}
During O1, the two aLIGO detectors operated from September 12, 2015 to January 19, 2016.
Ideally, we would like to compute the BCR for all triggers produced during this period to show that it can efficiently discriminate between glitches and CBC signals. However, computational limitations prevent this \footnote{There are $\mathcal{O}(10^7)$ background triggers in O1. The run time on a single background trigger using the \texttt{LALInference} implementation of nested-sampling is usually between $1$ to $5$ hours.}. 
Instead, we pick a subset of $\ntrigs$ multi-detector background binary-black-hole triggers identified by \texttt{PyCBC}, one of the staple search pipelines \cite{Nitz:2017svb,Usman:2015kfa,Canton:2014ena,pycbc-software}.
We pick the background triggers by sampling from the full trigger-set uniformly in the log of the inverse-FAR (${\rm IFAR}\equiv 1/{\rm FAR}$) for IFARs in $[\,5\times 10^{-5}, 10^{6}\,]$ yr, which is the total range reported by the pipeline.
This sampling allows us to analyze common (low IFAR) and rare (high IFAR) background events.

To compute the evidences making up the BCR, \eq{bcr}, we run the nested-sampling algorithm implemented in the \texttt{LALInference} library on 4s-long data segments containing each trigger \cite{Veitch:2014wba, TheLIGOScientific:2016wfe}.
Given the large number of triggers involved, this would not be feasible without the reduction in the computational cost of Bayesian inference provided by reduced order quadrature (ROQ) methods (see, e.g., \cite{Smith:2016qas}).
Using this technique makes no measurable difference for the values of the computed evidences.%
\footnote{For example, see Table IV in Appendix B of \cite{O1:BBH}, where Bayes factors computed with and without ROQ can be compared (the values in that example are close, but not identical due to differences in waveform approximants).}

Templates are produced using \texttt{IMRPhenomP}, a standard waveform family \cite{Husa:2015iqa,Khan:2015jqa, Hannam:2013oca, Smith:2016qas}.
We restrict the priors on the masses such that we only consider signals that are less than 4s in duration, resulting in a chirp-mass range of $12.3 M_{\odot} \leq \mathcal{M} \leq 44.7M_{\odot}$.
We further restrict the mass ratio to lie within $1 \leq q \leq 8$.
The dimensionless spin magnitudes are taken to be within $[0,0.89]$, and we consider all spin angles.
The prior on luminosity distance assigns probability uniformly in volume, with an upper cutoff of 5 Gpc.
These priors, as well as the priors for all other parameters, follow the default for standard \texttt{LALInference} analyses with ROQ \cite{Smith:2016qas, Veitch:2014wba}. 
The PSD used for matched filtering is calculated using the \texttt{BayesWave} algorithm \cite{2015PhRvD..91h4034L,2015CQGra..32m5012C}.

The search that originally produced our set of triggers considered a wider range of masses and spins than we do in the BCR computation for the purpose of this demonstration.
To accommodate this, we prescreened the background to only allow triggers with masses within our priors.
It would be straightforward in principle to broaden our constraints to encompass all triggers produced by the pipelines.
However, we refrain from doing so to keep our computational costs manageable.
Our preliminary analyses of slightly longer triggers (8s, 16s and 32s) yield results qualitatively similar to those presented below. 

We compare the BCRs from our background selection to several foreground triggers.
The foreground includes eight {\em hardware injections}, which were performed by physically actuating the test masses of the detectors to simulate signals similar to GW150914 \cite{2017PhRvD..95f2002B}.
We also analyze a set of \ninjs{} {\em software injections}: simulated signals inserted in O1 data, with arbitrary sky location and orientation, and with masses and spins that span our priors (in particular, the luminosity distance distribution is uniform in volume with cutoff at 2.5 Gpc).
On top of these artificial triggers, we also compute the BCR for GW150914 \cite{GW150914_paper} and LVT151012 \cite{O1:BBH}.
The freedom provided by the $\alpha$ and $\beta$ parameters in \eq{bcr} may be used to minimize the overlap between the simulated-signal and background distributions; the results below correspond to values of $\alpha=10^{-6}$ and $\beta = 10^{-4}$, but may be adjusted in future analyses (see Appendix \ref{app:weights}).

\begin{figure}
\includegraphics[width=\columnwidth]{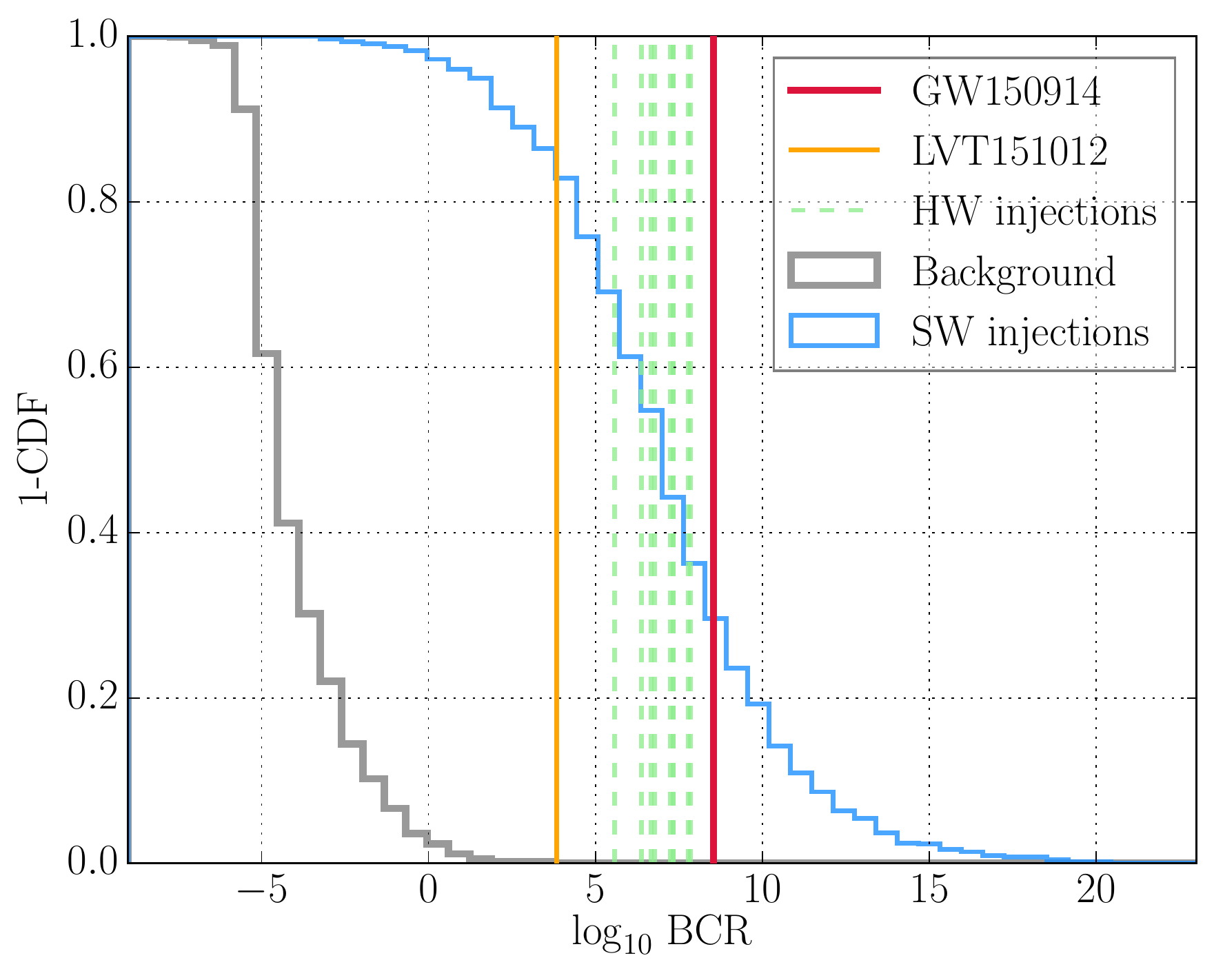}
\caption{{\em BCR distributions.} Histograms represent the survival function (1-CDF) from our selection of \ntrigs{} aLIGO O1 background triggers (gray), and \ninjs{} simulated signals (blue). Vertical lines mark the BCRs of eight hardware injections (dashed green), LVT151012 (leftmost, orange line), and GW150914 (thick red line). Background triggers were selected to be uniformly distributed in log-IFAR, and 98\% yield $\log{\rm BCR}<0$.
}
\label{fig:BCR_hist}
\end{figure}

\heading{\label{sec:results} Results}
\fig{BCR_hist} shows the BCR distributions obtained for background triggers and software injections. 
The figure also displays the values obtained for GW150914, LVT151012 and hardware injections, all of which show much stronger evidence for being coherent CBC signals, rather than incoherent glitches (high BCR).
We find a clear separation between injections and background events---suggesting that the BCR is good at distinguishing CBC signals from glitches.
If we consider the intrinsic probabilistic meaning of the BCR, a value of $\log {\rm BCR}<0$ indicates a preference for the instrumental-artifact hypothesis ($\hyp{I}$) over the coherent-signal one ($\hyp{S}$).
As expected, the vast majority (98\%) of background triggers fall bellow this mark, while the opposite is true for injections.
GW150914 and LVT151012 yield $\log_{10} {\rm BCR}$ values of 8.5 and 3.8 respectively.

\begin{figure}
\includegraphics[width=\columnwidth]{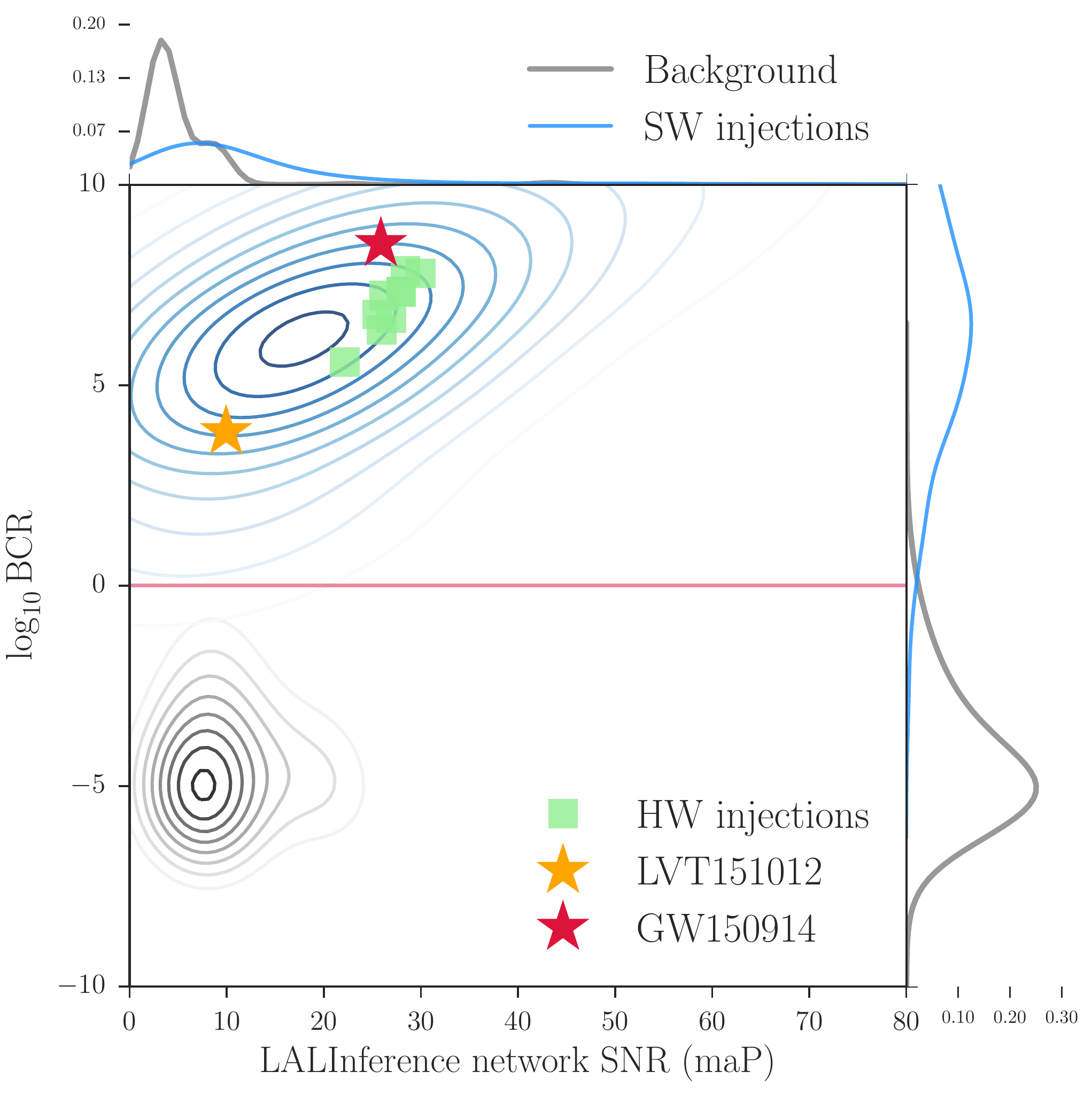}
\caption{{\em BCR vs SNR distributions.} Contours represent the normalized probability density of selected background triggers (gray) and simulated signals (blue) in log-BCR vs SNR space.
The plot also shows eight hardware injections (green squares), LVT151012 (orange star), and GW150914 (red star).
The curves shown on the right (top) result from a Gaussian kernel-density estimation of the one-dimensional distribution of log-BCRs (SNRs), obtained after integration over the $x$-axis ($y$-axis).
Background triggers were selected to be uniformly distributed in log-IFAR, and 98\% yield $\log{\rm BCR}<0$ (threshold marked by a horizontal red line for convenience).
The SNR on the $x$-axis is the coherent matched-filter signal-to-noise ratio of the template recovered with maximum {\em a posteriori} probability (maP) by our inference pipeline (\texttt{LALInference}).
}
\label{fig:BCR_SNR}
\end{figure}

\fig{BCR_SNR} shows the same populations from \fig{BCR_hist}, plotted also as a function of the network signal-to-noise ratio (SNR) recovered by our coherent Bayesian analysis.
\fig{BCR_SNR} reveals that the BCR values of the signal population are correlated with SNR, which reflects the fact that we are better able to evaluate the coherence of signals that stand clearly above the noise floor. 
As a result, the separation between our signal and glitch populations improves with SNR.
Because this population of background triggers was purposely selected to be uniform in log-IFAR, the gray contours in \fig{BCR_SNR} should not be taken to be representative of the actual glitch distribution: this would include {\em vastly} more low-SNR triggers.
In any case, BCR is largely independent of SNR for background triggers.

There are three software injections with ${\rm SNR} > 12$, but ${\rm BCR}<1$.
This is due to two characteristics that make the noise model preferable: $(i) $ the ratio of SNRs in two detectors is greater than three, and $(ii)$ the signal in at least one detector is too weak to be confidently discernible from Gaussian noise (${\rm SNR} \sim 5.5$).
These rare circumstances are caused by source locations and orientations unfavorable to the detector network, and, as such, should be mitigated by adding more instruments.

\begin{figure}
\includegraphics[width=\columnwidth,trim=0.5cm 0 0 -0.4cm]{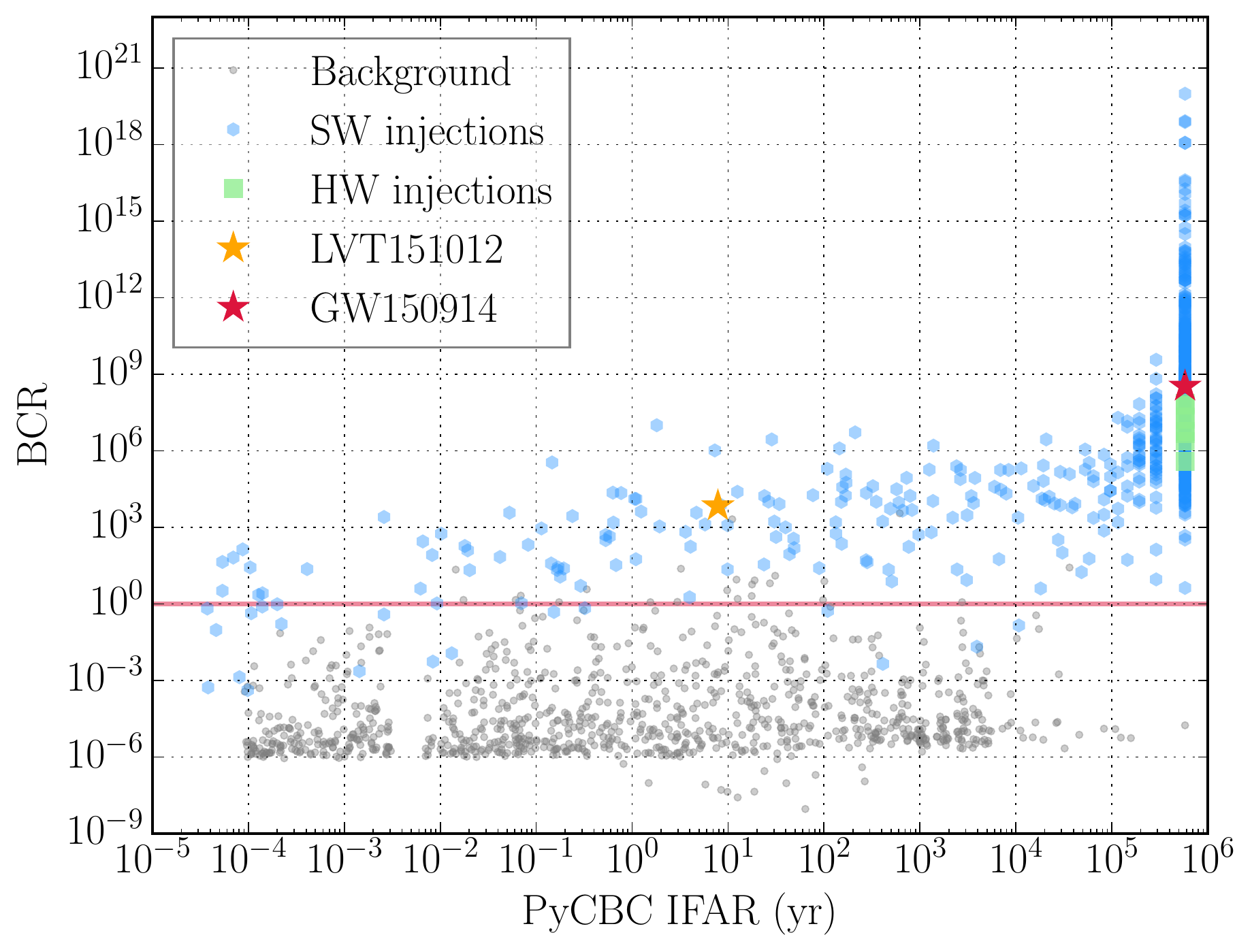}
\caption{{\em BCR vs IFAR.} BCR for the same data shown in \fig{BCR_hist}, plotted vs the inverse false-alarm-rate (IFAR) assigned to each event by \texttt{PyCBC}, one of the staple aLIGO search pipelines. There are six background triggers with ${\rm BCR} \ll 10^{-9}$ which fall outside the range of this plot; no foreground triggers are excluded from this plot. High-significance events pile up on the right because their IFAR is a lower limit determined by the most significant trigger in the background. This plot suggests the BCR may be used to more easily reject incoherent glitches.
}
\label{fig:BCR_IFAR}
\end{figure}

Irrespective of its Bayesian interpretation, we may treat the BCR as a traditional detection statistic to obtain a frequentist estimate of the significance of any given foreground event based on the measured background (e.g.~a $p$-value, or better, a likelihood ratio).
Again, our background triggers were selected to represent common and rare events in equal numbers, so the distribution in \fig{BCR_hist} need not be the same as that of the entire background, and should not be used for this purpose.
However, as shown in \fig{BCR_IFAR}, we find that there is no evidence for strong correlation between BCR and the IFAR assigned by the detection pipeline.
This suggests that the background BCR distribution shown in \fig{BCR_hist} is likely representative of the whole.
Furthermore, \fig{BCR_IFAR} implies that the BCR may be used to more easily reject incoherent glitches, irrespective of IFAR, and thus increase our detection confidence for marginal events like LVT151012.

\heading{\label{sec:implementation} Future implementation}
Given its ability to separate signals from glitches, the BCR may supplement existing search strategies and help increase their sensitivity, even with existing computational resources.
The most straightforward way to achieve this would be to run existing CBC pipelines as usual, with an extra threshold on BCR (e.g.~discarding any triggers with, say, ${\rm BCR}<1$).
Our results suggest that this would be an efficient way of discarding the vast majority of instrumental artifacts, thereby increasing detection confidence of real signals \cite{Kanner:2015xua}.

Computational costs would currently preclude obtaining BCRs for {\em all} triggers (foreground and background) produced during a regular observation run, so this extra step would have to be reserved for the most significant ones, as determined by the main pipeline.
However, processing all triggers \textit{would} have the added advantage of potentially enabling the detection of weak GW events that would otherwise be missed (e.g.~low-IFAR, but high-BCR, injections in \fig{BCR_IFAR}).
In the future, this would also enable us to move beyond a simple BCR veto, and instead use large numbers of simulated signals and background events to define empirical probability distributions over a space of multiple figures of merit (e.g.~BCR and SNR, as in \fig{BCR_SNR}).
This could be used to obtain likelihood ratios to categorize a trigger as signal or noise---which can be shown to be an optimal strategy for classification problems such as this, and have been used successfully by some existing searches \cite{Cannon2013FAR,gstlal-methods,Lynch:2015yin}.
Future improvements in ROQ methods, like their implementation on graphical processing units, will be vital in making this possible.

The values of the $\alpha$ and $\beta$ weights in \eq{bcr} have a strong effect on the shape of the distributions of \fig{BCR_SNR}, as discussed in Appendix \ref{app:weights}.
While here we have set them to values that yield a good separation between the signal and background populations, future studies may systematically optimize these parameters using a more comprehensive set of software injections and a large, representative set of background triggers.
This may be achieved via any standard optimization scheme that attempts to minimize the overlap between the two populations. 
The values would, of course, be fixed before analyzing any foreground data.

\heading{Conclusion}
We have demonstrated that Bayesian models based on the coherence of GW triggers across detectors may successfully distinguish between real CBC signals and transient instrumental noise (Figs.~\ref{fig:BCR_hist} and \ref{fig:BCR_SNR}).
We introduced a specific figure of merit, the BCR, which responds to incoherent glitches in a way that is complementary to that of standard CBC pipelines (\fig{BCR_IFAR}).
Finally, we suggested a few avenues for incorporating this (or similar) measure of coherence into existing GW search strategies, the simplest of which would take the form of a new veto for detection candidates.
This could be implemented today to increase the number of gravitational waves confidently detected by LIGO and Virgo, without needing to further improve detector hardware.

Versions of the ranking statistic used by \texttt{PyCBC} in recent analyses have incorporated some measure of coherence \cite{Nitz:2017svb}, and it remains to be seen whether this introduces some correlation between BCR and IFAR in \fig{BCR_IFAR}.
Furthermore, while this study focused on detection candidates produced by the two aLIGO detectors during O1, we are currently investigating how the power of the BCR is affected by the addition of new detectors, like Virgo.
Finally, although here we focused on short-duration (4s) triggers from high-mass binary-black-hole mergers, our preliminary results on slightly longer (8s, 16s and 32s) show qualitatively similar behavior.

\begin{acknowledgments}
We thank Alan Weinstein, Alex Nitz, Carl-Johan Haster, Stefan Hild, Reed Essick, Ryan Lynch, Colm Talbot, Eric Thrane, John Veitch and Thomas Dent  for helpful comments.
Rory Smith is supported by the Australian Research Council
Centre of Excellence for Gravitational Wave Discovery
(OzGrav), through project number CE170100004.
%
The authors thank the LIGO Scientific Collaboration for access to the data and gratefully acknowledge the support of the United States National Science Foundation (NSF) for the construction and operation of the LIGO Laboratory and Advanced LIGO (PHY-0757058), as well as the Science and Technology Facilities Council (STFC) of the United Kingdom, and the Max-Planck-Society (MPS) for support of the construction of Advanced LIGO. Additional support for Advanced LIGO was provided by the Australian Research Council.
This manuscript has LIGO Document ID \dcc{}.
\end{acknowledgments}

\appendix
\section{Effect of BCR weights} \label{app:weights}
\newcommand{\alphamain}{10^{-6}}
\newcommand{\betamain}{10^{-4}}

The weights $(\alpha, \beta)$ that go into the calculation of the BCR in \eq{bcr} have a critical impact on the  degree of separation between the signal and glitch populations.
Here we elaborate on this point, and show how we improve upon previous work by explicitly taking advantage of the extra freedom afforded by these parameters.

From a Bayesian perspective, $\alpha$ and $\beta$ encode our prior beliefs on the relative probabilities of each of the submodels that are compared in the computation of the BCR:
$\alpha$ determines by what factor the coherent-signal hypothesis ($\hyp{S}$) should be favored over the instrumental-feature hypothesis ($\hyp{I}$),
\begin{equation} \label{e:alpha}
\alpha \equiv \frac{P(\hyp{S})}{P(\hyp{I})}\, ,
\end{equation}
while $\beta$ gives the probability of the glitch hypothesis ($\hyp{G}$) conditional on the assumption that there is an instrumental-feature to begin with,
\begin{equation} \label{e:beta}
\beta \equiv P({\hyp{G}}_i \mid \hyp{I}) = 1 - P({\hyp{N}}_i \mid \hyp{I})\, ,
\end{equation} 
for any dectector $i$, as discussed in Sec.~\ref{sec:bcr}. 
The last equality in \eq{beta} uses the fact that we {\em define} the instrumental-feature hypothesis as the logical union of the glitch and Gaussian noise ($\hyp{N}$) subhypotheses, i.e.~$\hyp{I} \equiv \hyp{G} \lor \hyp{N}$, and that the latter are logically disjoint, i.e.~$\hyp{G} \land \hyp{N}={\rm False}$, so $P(\hyp{N} \mid \hyp{G}) = P(\hyp{G} \mid \hyp{N}) = 0$.

\begin{figure}
\includegraphics[width=\columnwidth]{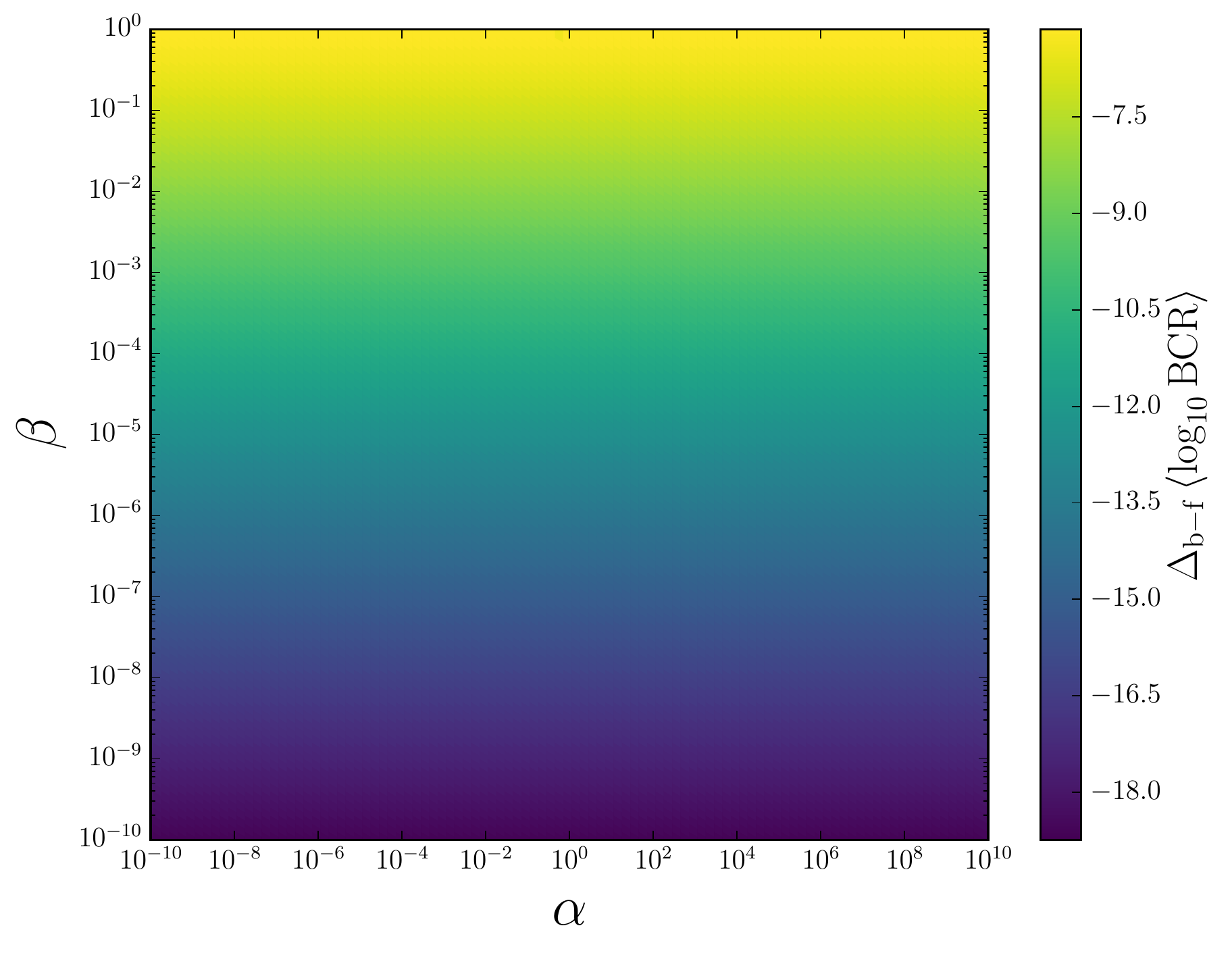}
\caption{{\em Effect of weight on population separation.}
Color represents the difference in mean $\log{\rm BCR}$ between background and foreground, $\Delta_{\rm b-f} \left\langle \log {\rm BCR} \right \rangle$ as defined in \eq{delta_bf}.
This is shown as a function of the BCR prior weights, $\alpha$ ($x$-axis) and $\beta$ ($y$-axis), of \eq{bcr}.
All values are negative because the foreground always has larger mean, so darker colors correspond to greater distance between the population means. 
}
\label{fig:weights_bcr-mean-difs}
\end{figure}

It follows from the probabilistic interpretation of these parameters that their allowed ranges are $0<\alpha<\infty$ and $0\leq\beta\leq 1$.
All results presented in the main text were produced using the values
\begin{equation} \label{e:mainweights}
\text{Main text:} ~~ \left(\alpha = \alphamain, \beta = \betamain \right) .
\end{equation}
This specific choice was made to yield a good separation between the background and foreground populations, as reflected by Figs.~\ref{fig:BCR_hist} and \ref{fig:BCR_SNR}.
These values also result in an overall normalization such that ${\rm BCR}=1$ gives the point at which both hypotheses are equally likely given {\em our} trigger set (i.e.~the horizontal red line in \fig{BCR_SNR} roughly agrees with the intersection of the blue and gray curves on the right panel).

\begin{figure}
\includegraphics[width=\columnwidth]{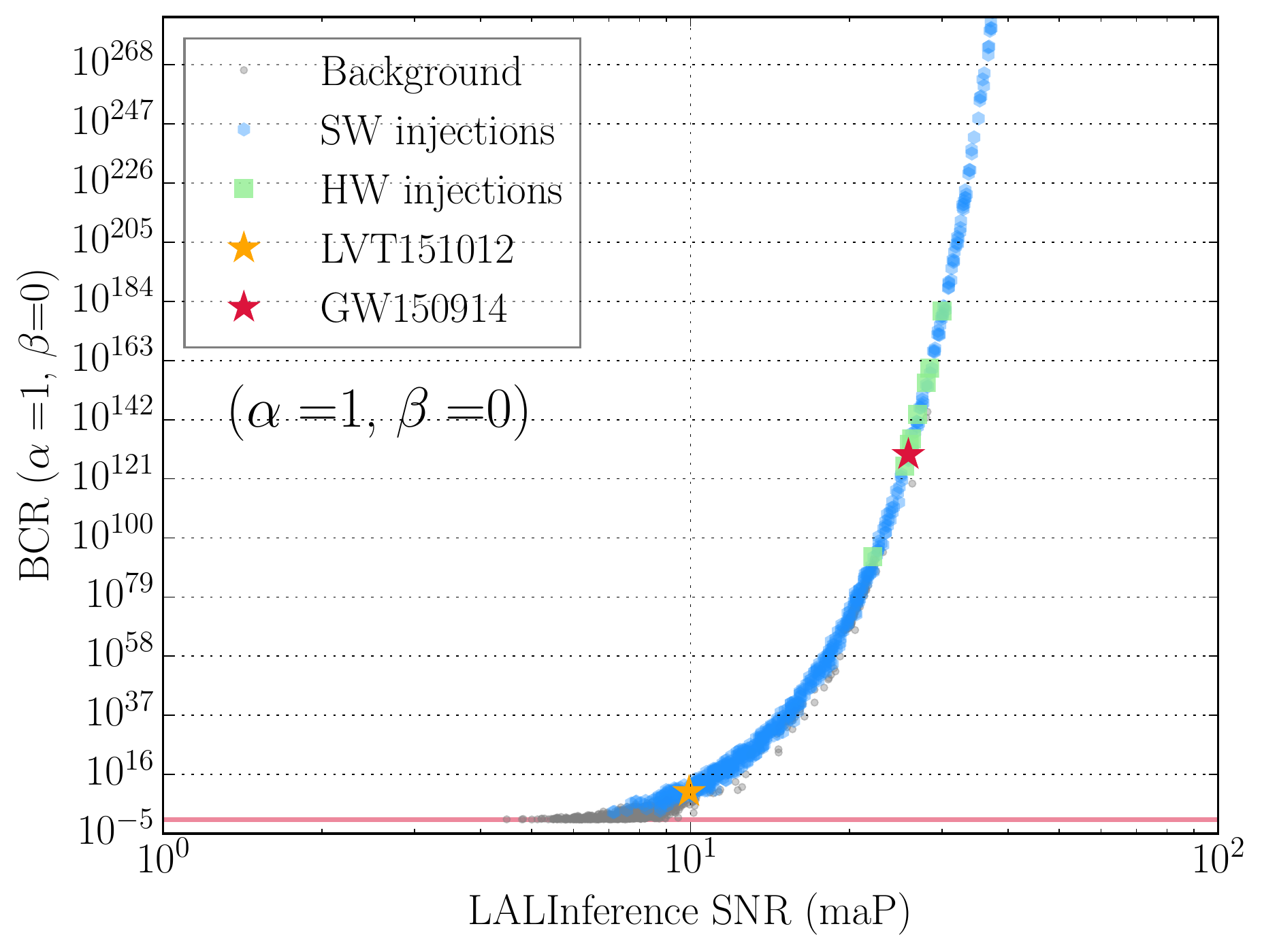}
\caption{{\em BCR $(\alpha=1, \beta=0)$ vs SNR.} BCR vs SNR for the same data shown in Figs.~\ref{fig:BCR_hist}--\ref{fig:BCR_IFAR}, but with analyzed with $(\alpha=1, \beta=0)$.
For this choice of weights, the BCR reduces to the Bayesian odds between signal and Gaussian noise, \eq{bsn}, and scales with SNR according to \eq{bsn_snr}, for both background (gray circles) and foreground (blue hexagons).
The SNR on the $x$-axis is the coherent matched-filter signal-to-noise ratio of the template recovered with maximum {\em a posteriori} probability (maP) by our inference pipeline (\texttt{LALInference}).
}
\label{fig:weights_nobeta}
\end{figure}

To see how $\alpha$ and $\beta$ impact the separation between the background and foreground populations, consider as a proxy the distance between the mean BCRs for the two populations.
In particular, define the quantity
\begin{equation} \label{e:delta_bf}
\Delta_{\rm b-f} \left\langle \log {\rm BCR} \right \rangle \equiv
\left\langle \log {\rm BCR}^{\rm (b)}\right\rangle - \left\langle \log {\rm BCR}^{\rm (f)} \right \rangle ,
\end{equation}
where the angle brackets on the right denote averaging over triggers, and the superscripts ``(b)'' and ``(f)'' refer to ``background'' and ``foreground'' respectively.
This number then gives a measure of the vertical distance between the centers of the distributions in \fig{BCR_SNR}.
The effect of $\alpha$ and $\beta$ on this quantity is shown in \fig{weights_bcr-mean-difs}, where darker colors correspond to greater absolute mean distance.
As expected from \eq{bcr}, the separation is a strong function of $\beta$, while it is largely independent of $\alpha$.
It can also be seen from \eq{bcr} that $\alpha$ should merely impact the overall normalization of the BCR, shifting all values up or down.

By tuning $\beta$ we may thus control the degree of bias introduced in the computation of the BCR.
This can be used to correct for shortcomings in the definitions of the noise submodels themselves, so as to best distinguish foreground and background. 
The reason this is necessary in the first place is that not all glitches will conform strictly to the ``worst-glitch'' hypothesis as we have defined it via \eq{zg}.
For instance, the distribution of glitch morphologies and SNRs need not conform to the parameter priors assumed in the computation of $\evidence^{G}$; instead of tuning the parameter priors, one may correct for this effect via $\beta$ (which is easier to implement).

Looking at \fig{weights_bcr-mean-difs}, one may be tempted to substantially reduce $\beta$ to maximize the distance between the distribution means.
However, the quantity plotted in \fig{weights_bcr-mean-difs}, \eq{delta_bf}, is insensitive to the fact that the two distributions do not retain their shape when $\beta$ is varied, and therefore is only useful as a proxy for population overlap when looking at small changes in the weights.
In other words, \fig{weights_bcr-mean-difs} fails to convey the fact that there is a {\em penalty} in introducing too strong of a bias through $\beta$.
This is related to the {\em bias-variance tradeoff}, well known in statistical inference (see e.g.~\cite{Hastie2009}).
Let us explore how this tradeoff is manifested throughout the range of valid values for $\beta$. 

\begin{figure}
\includegraphics[width=\columnwidth]{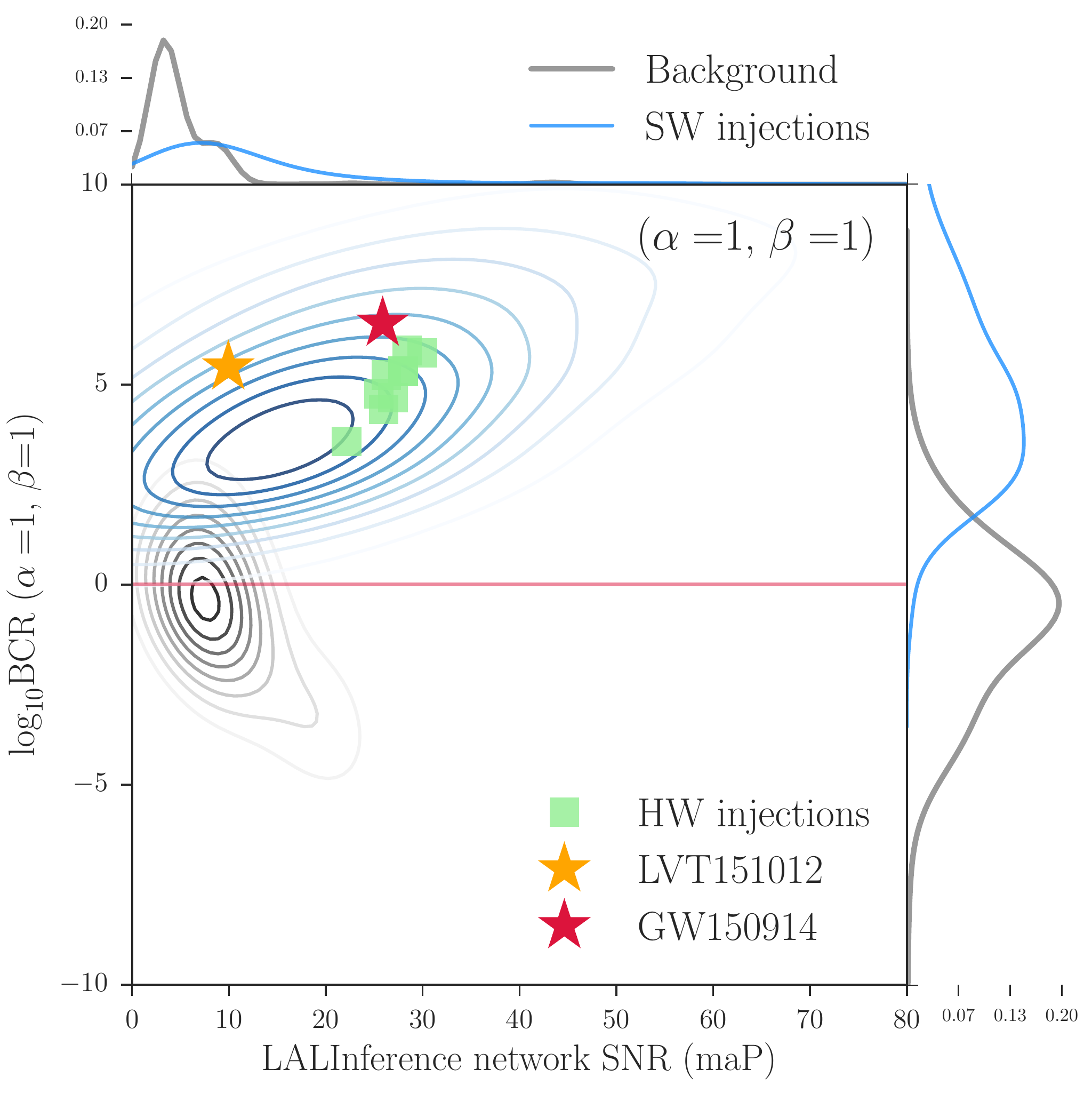}
\caption{{\em BCR $(\alpha=1, \beta=1)$ vs SNR distributions.} 
This plot is completely analogous to \fig{BCR_SNR}, but with $(\alpha=1, \beta=1)$ instead of $(\alpha=\alphamain, \beta=\betamain)$ [cf.~\eq{bcr}].
For this choice of weights, the BCR reduces to the BCI, \eq{bci}, resulting in greater overlap between the background (gray) and foreground (blue) distributions.
For more details about this plot, refer to the caption of \fig{BCR_SNR}.
}
\label{fig:weights_BCR_SNR_beta1}
\end{figure}

On one end, setting $\beta=0$ comes at the price of throwing away all information about the incoherence of the trigger.
As can be deduced from \eq{bcr}, in the limit of vanishing $\beta$ the BCR is nothing but the usual signal vs Gaussian-noise odds (BSN), 
\begin{equation} \label{e:bsn}
{\rm BCR}(\alpha=1, \beta=0) = \evidence^{S} / \evidence^{N} \equiv {\rm BSN}\, , 
\end{equation}
and the glitch model is totally ignored. 
For this choice of $\beta$, the BCR will just follow the usual dependence of BSN on SNR (see, e.g., \cite{Creighton2011}),
\begin{equation} \label{e:bsn_snr}
\log {\rm BSN} \propto {\rm SNR}^2\, ,
\end{equation}
irrespective of whether the trigger is a glitch or a coherent signal, as shown in \fig{weights_nobeta}.
Although the distance between the means of the two populations in this figure is large (as reflected also by \fig{weights_bcr-mean-difs}, for $\beta\rightarrow 0$), this is only because, on average, the background triggers in our set have lower SNR than the foreground.

On the other end, setting $\beta=1$ is equivalent to ignoring the possibility that the trigger was produced by Gaussian noise.
In that case, the BCR reduces to the evidence ratio between the coherent-signal and incoherent-glitch hypotheses, a quantity often called ``BCI'' by gravitational-wave data analysts (assuming $\alpha=1$):
\begin{equation} \label{e:bci}
{\rm BCR}(\alpha=1, \beta=1) = \evidence^{S} / \evidence^{G} \equiv {\rm BCI}\, .
\end{equation}
The use of this quantity for glitch-discrimination purposes in CBC searches was proposed in \cite{VV2010}.
However, we find that it does not produce a sufficient separation between the background and foreground populations, except for loud triggers.
For example, while $(\alpha=\alphamain, \beta=\betamain)$ yields \fig{BCR_SNR}, $(\alpha=1, \beta=1)$ yields \fig{weights_BCR_SNR_beta1}.
From this plot, it is easy to see that the BCI is good at distinguishing {\em loud} incoherent glitches from {\em loud} coherent signals, but is inconclusive for weak triggers.

\begin{figure}
\includegraphics[width=\columnwidth,trim=0.5cm 0 0 -0.4cm]{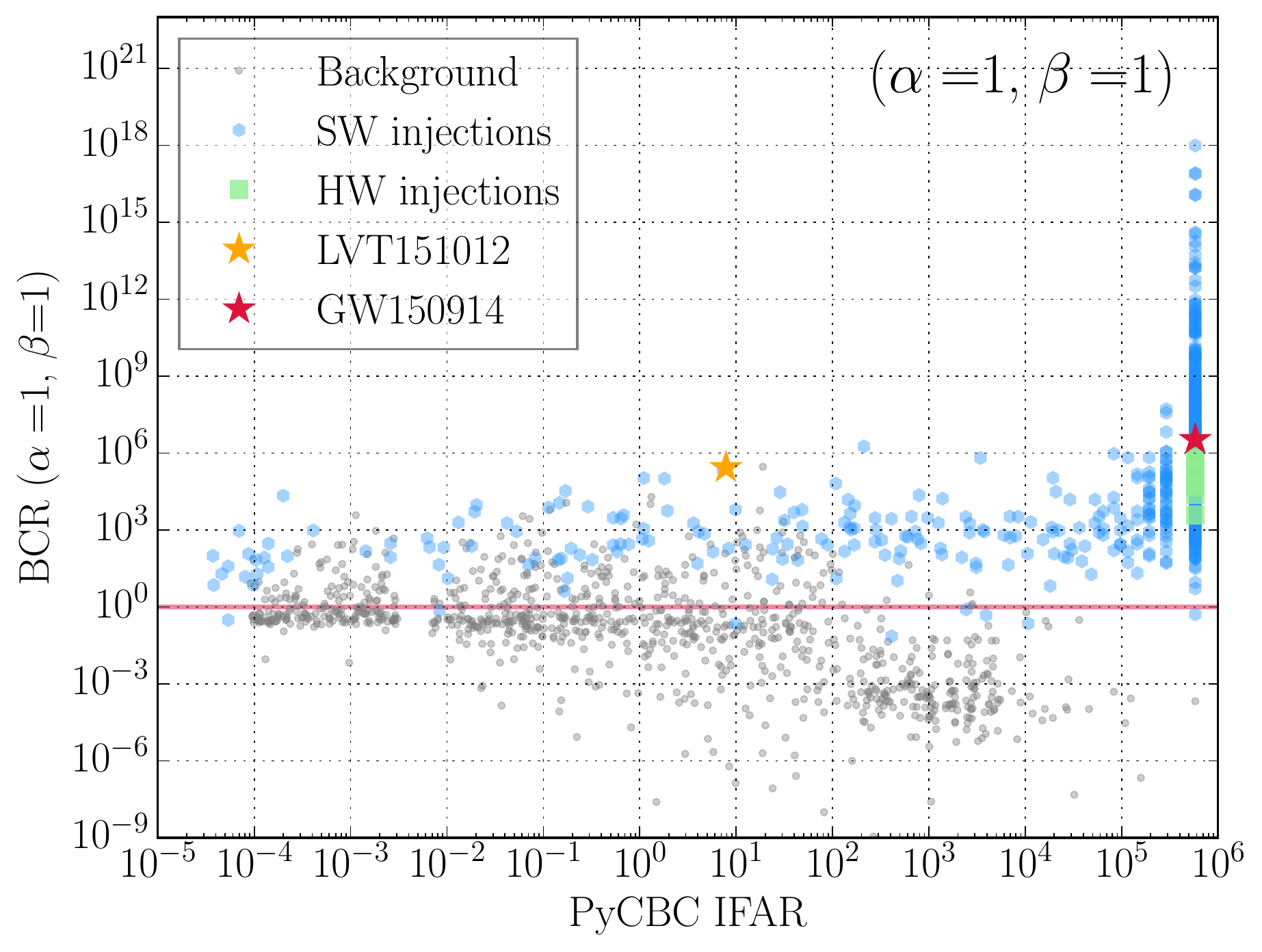}
\caption{{\em BCR $(\alpha=1, \beta=1)$ vs IFAR.}
This plot is completely analogous to \fig{BCR_IFAR}, but with $(\alpha=1, \beta=1)$ instead of $(\alpha=\alphamain, \beta=\betamain)$ [cf.~\eq{bcr}].
For this choice of weights, the BCR reduces to the BCI, \eq{bci}, resulting in greater overlap between the background (gray) and foreground (blue) distributions.
For more details about this plot, refer to the caption of \fig{BCR_IFAR}.
}
\label{fig:weights_BCR_IFAR_beta1}
\end{figure}

We can check that changing $\beta$ indeed affects primarily {\em weak} glitches by comparing \fig{weights_BCR_IFAR_beta1} to \fig{BCR_IFAR}, BCR vs IFAR plots which were produced with $\beta=1$ and $\beta=\betamain$ respectively.
The change in $\beta$ from \fig{weights_BCR_IFAR_beta1} to \fig{BCR_IFAR} causes low-IFAR (low-SNR) glitches to yield significantly lower BCRs, while high-IFAR (high-SNR) triggers are largely unaffected.
Importantly, low-IFAR (low-SNR) signals are also down-ranked after the change, but to a lesser degree on average
Hence the separation in BCR improves, as can be seen by comparing the right panels of \fig{weights_BCR_SNR_beta1} and \fig{BCR_SNR}.

\begin{figure}
\includegraphics[width=\columnwidth]{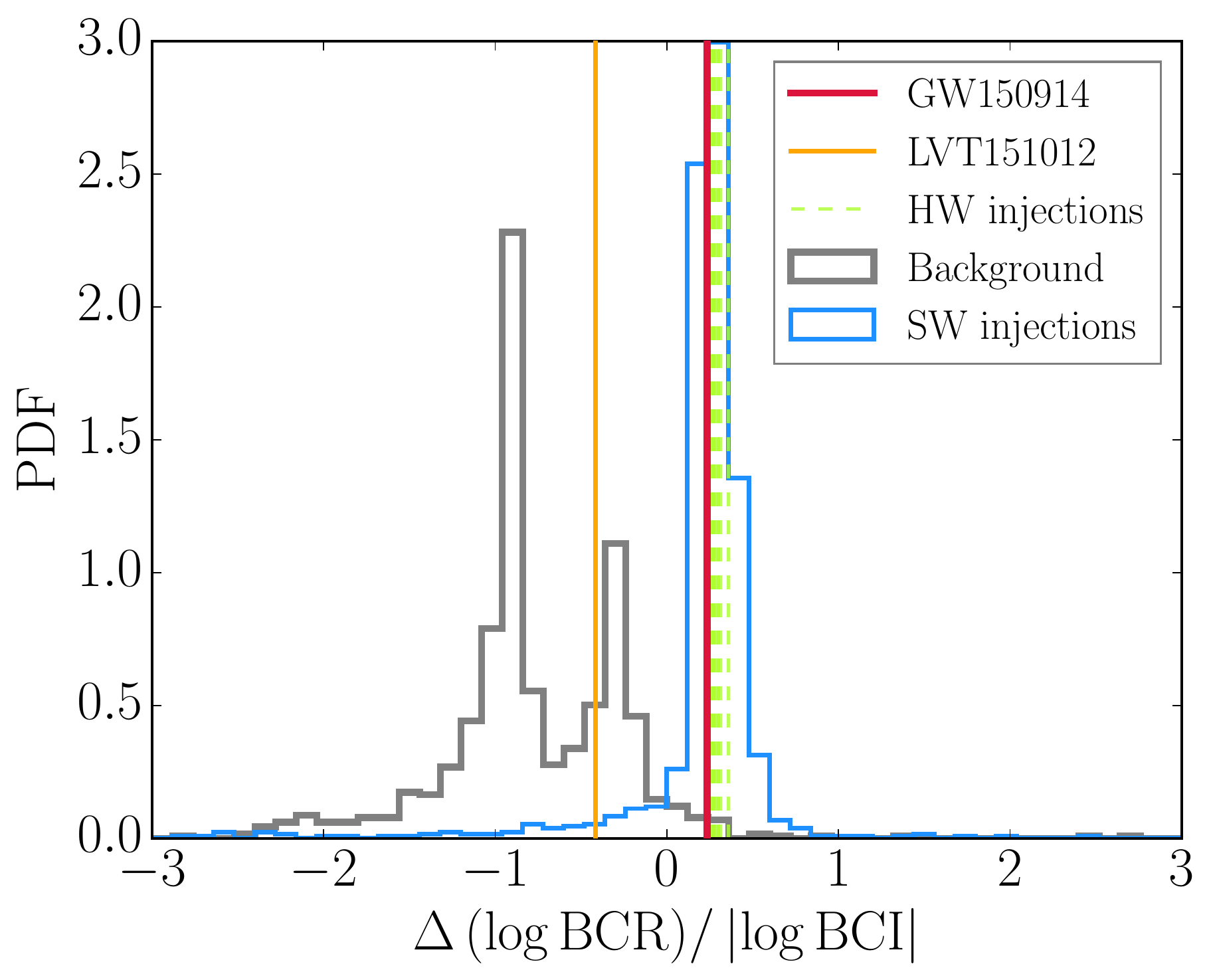}
\caption{{\em Effect of weights on $\log$BCR.}
Histogram of the fractional change in $\log{\rm BCR}$ when going from $(\alpha=1, \beta=1)$ to $(\alpha=\alphamain, \beta=\betamain)$, \eq{delta_bci}.
This plot summarizes the differences between the BCRs shown in Figs.\ \ref{fig:BCR_SNR} \& \ref{fig:BCR_IFAR} and those in Figs.\ \ref{fig:weights_BCR_SNR_beta1} \& \ref{fig:weights_BCR_IFAR_beta1}.
}
\label{fig:weights_change_hist}
\end{figure}

To further quantify the effect of $\beta$, we can also look at the fractional change in $\log{\rm BCR}$ when going from $(\alpha=1, \beta=1)$ to $(\alpha=\alphamain, \beta=\betamain)$,
\begin{equation} \label{e:delta_bci}
\frac{\Delta \left( \log {\rm BCR}\right)}{|\log{\rm BCI}|} \equiv \frac{\log{\rm BCR} {(\alphamain, \betamain)} - \log{\rm BCI}}{|\log {\rm BCI}|}\, ,
\end{equation}
where vertical bars mark absolute values, and the BCI is defined by \eq{bci}.
This quantity is histogrammed in \fig{weights_change_hist} for the triggers in our set.
The fact that the change in $\beta$ affects weak glitches more significantly than strong ones is reflected in the bimodality of the gray distribution: the left (right) peak corresponds to triggers below (above) an effective threshold of ${\rm SNR} \sim 9$.
On the other hand, the blue distribution in \fig{weights_change_hist} shows that most (although not all) signals are largely unaffected by the change in $\beta$, with a mean increase in BCR but long tails extending mainly to the left.
This large variance is due mostly to the weaker signals for which the BCR decreased due to the change in $\beta$.

By tuning the weights, we may attempt to find a sweet-spot in which the bias introduced is just enough to separate weak glitches from weak signals, without confounding loud glitches with loud signals.
The choice of \eq{mainweights} was found to be close to this ideal, and achieves this by separating the weak glitches in our set from the weak signals to an extent, largely without altering loud triggers (Figs.\ \ref{fig:BCR_hist}--\ref{fig:BCR_IFAR}).

\bibliography{refs,cbc-group}

\end{document}